%% file: BDMS_ICPAD2018_8p.tex
\documentclass[conference]{IEEEtranNew}
\IEEEoverridecommandlockouts
\usepackage{graphicx,epsfig,subfigure,footnote}
\usepackage{cite}
\def\BibTeX{{\rm B\kern-.05em{\sc i\kern-.025em b}\kern-.08em
    T\kern-.1667em\lower.7ex\hbox{E}\kern-.125emX}}
\begin{document}

\title{biggy: An Implementation of Unified Framework for Big Data Management System}

\author{
\IEEEauthorblockN{Yao Wu, Henan Guan}
\IEEEauthorblockA{Key Laboratory of Data Engineering and Knowledge Engineering of Ministry of Education, Beijing, China\\
School of Information, Renmin University of China, Beijing, China\\
Email: \{ideamaxwu, henanguan\}@ruc.edu.cn}
}

\maketitle

%ABSTRACT
\begin{abstract}
Various tools, softwares and systems are proposed and implemented to tackle the challenges in big data on different emphases, e.g., data analysis, data transaction, data query, data storage, data visualization, data privacy. In this paper, we propose datar, a new prospective and unified framework for Big Data Management System (BDMS) from the point of system architecture by leveraging ideas from mainstream computer structure. We introduce five key components of datar by reviewing the current status of BDMS. Datar features with configuration chain of pluggable engines, automatic dataflow on job pipelines, intelligent self-driving system management and interactive user interfaces. Moreover, we present biggy as an implementation of datar with manipulation details demonstrated by four running examples. Evaluations on efficiency and scalability are carried out to show the performance. Our work argues that the envisioned datar is a feasible solution to the unified framework of BDMS, which can manage big data pluggablly, automatically and intelligently with specific functionalities, where specific functionalities refer to input, storage, computation, control and output of big data. 
\end{abstract}


%INTRO
\section{Introduction}

\subsection{Motivation of Datar}
As Alan Turing proposed the question ``Can machines think?'' \cite{turing1950computing}, the imitation game begins. Von Neumann started an engineering research on computer and described a logical design of a computer using the stored-program concept, which is known as the Von Neumann architecture \cite{von1993first}. Charles Babbage proposed the Analytical Engine, a designed mechanical general-purpose computer. The goal of these pioneers is to design a computing machine better than human brains, which can liberate human from manual work and tedious computation.

During the last several decades, data management principles such as relational model of data, physical and logical independence, declarative querying and cost-based optimization have led to several fields of researches and a prosperous industry. Many novel challenges and opportunities associated with big data necessitate rethinking many aspects of these data management platforms, while retaining other desirable aspects. The practice and theory contributions \cite{Bachman66,Codd70} of Bachman and Codd open up the research on database. And the steps on the road to data management never stop such as, Ingres \cite{StonebrakerWKH76}, Postgres \cite{Stonebraker90}, Mariposa \cite{StonebrakerALPSSSY96}, C-Store \cite{StonebrakerABCCFLLMOORTZ05}, VoltDB \cite{StonebrakerW13}, AsterixDB\cite{alsubaiee2014asterixdb} and P-Store \cite{TaftESLASMA18} in database field, as well as, Megastore \cite{Baker2011Megastore}, Spanner \cite{Corbett2013Spanner}, MillWheel \cite{Akidau2013MillWheel}, Azure CosmosDB\footnote{https://azure.microsoft.com/zh-cn/services/cosmos-db/} and TiDB\footnote{https://www.pingcap.com/} in distributed system field.

Michael Stonebraker proposed ``On Size Doesn't Fit All'', and in this paper, we try to argue that ``All Can Fit in One''. Since the computing power of machines becomes stronger, we can sniff the shift from computation to data management to explore more in-sight information and knowledge from data. Jim Gray foresighted the transformation from computation-intensive to data-intensive science discovery and brought forward The Fourth Paradigm \cite{Hey2012The}. He also thought the way to cope with such paradigm was to develop a new generation of computing tools to manage, visualize, and analyze massive data. As we all know, Big Data Management System (BDMS) is a complex set of functionalities, we think it necessary to propose a unified architecture to guide the design of BDMS. From these observations, we summarize and conclude with five main components in BDMS to provide a better explanation for a full understanding of our proposed datar architecture.

\begin{figure}
\centering
\includegraphics[width=1.0\columnwidth]{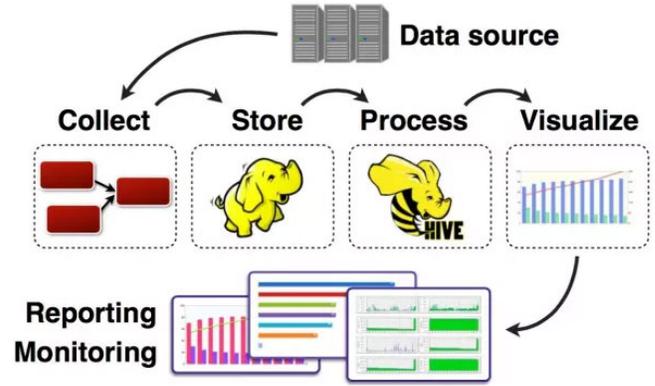}
\centering\caption{\label{flow}\footnotesize{A typical workflow for big data management.}}
\end{figure}

As shown in Fig. \ref{flow}, BDMS consists of several core components such as, collect, storage, process and visualize. Compared with traditional database systems, BDMS architecture is more flexible and open for varied requirements due to different focus-ons. In this paper, we unify the BDMS as \textbf{datar}, a general framework to design and build BDMS, corresponding to term \textbf{computer}.

\begin{figure*}
\subfigure[Computer Architecture]{
    \includegraphics[width=0.9\columnwidth]{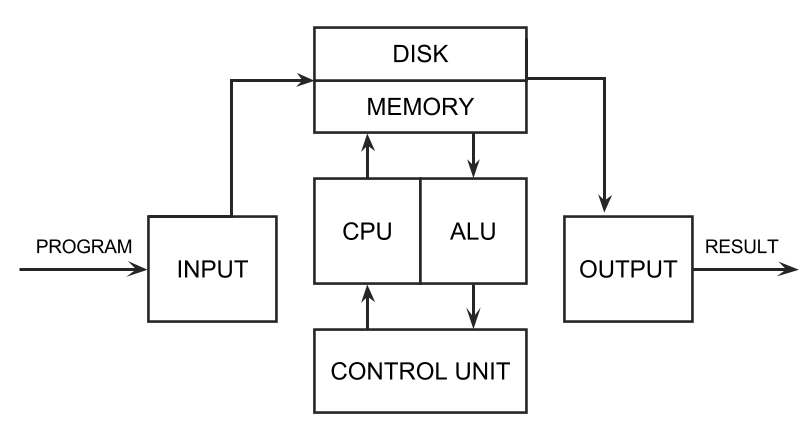}}
\subfigure[Datar Architecture]{
    \includegraphics[width=0.9\columnwidth]{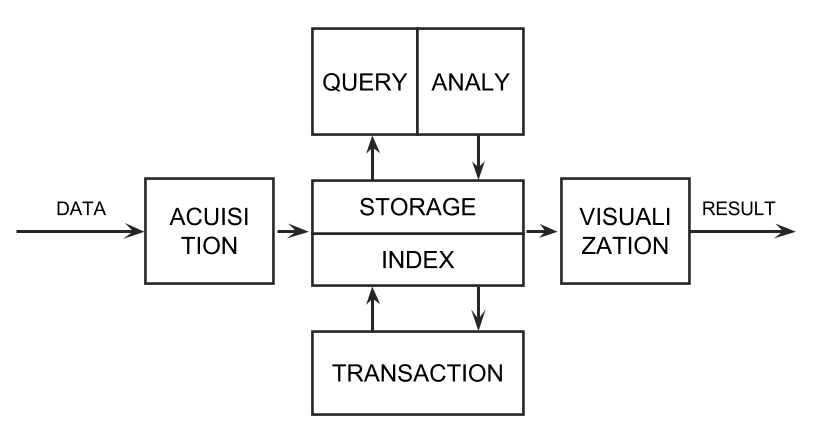}}
\centering\caption{\label{comp}\footnotesize{Computer VS Datar comparison in terms of architecture.}}
\end{figure*}

As we all know, the mainstream computer architecture is divided into five parts, i.e., input, storage, computation, control and output, in which, computation is the center. If we look closely, we can find that, BDMS is much the same as computer, consisting of (data) input, (data) storage, (data) computation (query/analysis), (data) control (transaction/recovery) and data output (visualization), in which, data storage is the center. In other words, we can name a computer Fast Computation  Processing System (FCPS). Likewise,  the BDMS can be called as datar, focusing on data. We use Fig. \ref{comp} to illustrate the similarities and differences between a computer and a datar, in terms of architecture. In Fig. \ref{comp} (a), five core components of a computer are shown in separate rectangles, while in Fig. \ref{comp} (b), five corresponding parts are shown. A computer and a datar share the similar functionalities with different emphases on computation or data storage.

\subsection{Concept of Datar}

\noindent \textbf{Definition (Datar)} \emph{A datar is a set of coherent softewares/systems based on a unified architecture that can manage (big) data pluggablly, automatically and intelligently with specific functionalities, where specific functionalities refer to input, storage, computation, control and output of the (big) data.} Datar is featured with Interactive Interface Clients, Pluggable Engines Configuration, Automatic Dataflow on Job Pipelines and Intelligent Self-driving System Management based on the unified framework. In this paper, we implement datar with these features as \textbf{biggy}, a data-storage-centered solution to datar implementation. 

A datar, i.e., a full-function BDMS, consists of five parts, data input, data storage, data computation, data control and data output. Compared with the computation-centered computer, a datar is data-centered. We take AsterixDB \cite{alsubaiee2014asterixdb} for example, which is a new, full-function BDMS. Data input is how data gets into the system. In AsterixDB, data feed is a built-in mechanism allowing new data to be continuously ingested into system from external sources, incrementally populating the datasets and their associated indexes \cite{grover2015data}. Data storage is how the data is stored in the system and how the indexes are built. In AstrixDB, data and index are stored based on LSM structure \cite{alsubaiee2014storage}. Data computation is how to mine valuable information from stored data. A bunch of methods can be applied, such as popular in-memory computation framework Spark on AsterixDB\cite{Bu2016Large}. Besides, the execution of data processing is also part of data analysis, like Hyracks \cite{borkar2011hyracks} in AsterixDB. Data control is how to control data when it is processed. It is different from the traditional database systems which have strict ACID properties. Another important aspect of datar is data output, e.g., visualization. Cloudberry\footnote{http://cloudberry.ics.uci.edu/} is a research prototype to support interactive analytics and visualization of large amounts of spatial-temporal data using AsterixDB. Based on these features of AsterixDB, it is ideal for us to explain the five main components of BDMS by one system. The key drawback of taking AsterixDB as BDMS is that it is a strongly coupled system, which is not suitable for varied and dynamic requirement in real scenarios when processing big data. And it is not easy for developers to combo it with new emerging engines. Datar is proposed to achieve a unified framework for building your own BDMS more flexible.

\subsection{Contributions of Datar}
With the development of Internet services, data contents are rapidly growing, and we have to face the challenges of handling such big data. Data system research has come into a new era, which brings the traditional concepts from row-based store to column-based store, from disk-based query to in-memory based analysis, and from ACID properties to CAP theorem. Big data shows great value in real application and challenges arise. Various tools and systems are proposed and developed to tackle these challenges on different emphases. In this paper, we describe the BDMS from a new perspective, the view of a computer architecture, to propose a unified framework datar. We focus our attention on the system architecture in BDMS and break it down into five main components to elaborate. The envisioned datar is implemented as biggy with favorable features. The key contributions can be summarized as,

\begin{itemize}
\item We review current big data management systems by five core components and state our contributions.
\item A unified architecture for big data management, i.e., datar, is proposed and explained to manage big data pluggablly, automatically and intelligently.
\item We propose ConfChain to connect the pluggable engines, a new data structure BigData to manage data flow, and Job Pipeline to execute jobs. 
\item We implement the envisioned architecture as biggy based on several popular engines to fulfill the functions of input, storage, computation, control and output, and demonstrate biggy in details by running examples.
\item We evaluate the performance of our proposed unified framework from different aspects to show the feasibility and potentials.
\end{itemize}

Rest of the paper is organized as follows. In Section \ref{bdbdms}, we introduce BDMS as datar from five aspects, input, storage, control, computation and output. Section \ref{bud} introduces the framework and key features of datar. The implementation of biggy and demonstrations are given in Section \ref{play}. Performance evaluation is carried out in Section \ref{eval}. Finally, we conclude with futures in section \ref{conc}.

\section{Break Down BDMS}\label{bdbdms}

%INPUT
\subsection{Data Input}\label{input}
Data can be input into storage since its generation or be ``inserted'' from other resources.

\subsubsection{Data Generation}

Big data can be generated from various sources, such as unstructured web data, enterprise internal data, government data, and other data from more sources, e.g., scientific applications data and pervasive sensing data. These datasets have their unique data characteristics in scale, time dimension, and data category.

\subsubsection{Data Feed}
Data feed is also known as data acquisition, having continuous data arrive into DBMS from external sources and incrementally populate a persisted dataset and associated indexes. In AsterixDB, a fault-tolerant data feed facility is built that scales through partitioned parallelism by using a high-level language. Apache Flume and Apache Kafka are two popular tools used as data feeds. 

%STORAGE
\subsection{Data Storage}
Data storage focuses on raw data storage and indexes storage in this paper. Besides, schemes storage, configuration files storage and views storage are also data storage but out of our scope.

\subsubsection{Data Storage}
Big data storage refers to the storage and management of large-scale datasets while achieving reliability and availability of data accessing. Various storage systems emerge to meet the demands of massive data. There are four main types of NoSQL databases: key-value DB, column-oriented DB, document-oriented DB and graph-oriented DB. Table \ref{popstore} shows examples of popular big data storage systems. 

\subsubsection{Data Index}
Index is an effective method to reduce the expense of disk I/Os, and improve query speeds. However, index has the additional cost for storing index files which should be maintained dynamically when data is updated. Basic structures include Hash table, Tree-based index, Multidimensional index, and Bitmap index. Big data index \cite{adamu2015survey} has additional requirements, such as parallelism and easily partitioned into pieces for parallel processing, e.g., Latent Semantic Indexing and HMM Indexing.

\begin{savenotes}
\begin{table}
\centering
\begin{tabular}{|l|l|l|l|}
\hline
\textbf{Key-Value} &\textbf{Column} &\textbf{Document} &\textbf{Graph} \\
\hline
\hline
Dynamo \cite{decandia2007dynamo} &BigTable \cite{chang2008bigtable} &MongoDB \cite{banker2011mongodb} &Giraph \footnote{http://giraph.apache.org/}\\
\hline
Voldemort \cite{sumbaly2012serving} &Cassandra \cite{lakshman2010cassandra} &SimpleDB\footnote{https://aws.amazon.com/simpledb/} &Neo4j \footnote{https://neo4j.com/}\\
\hline
Redis \cite{carlson2013redis} &HBase \cite{george2011hbase} &CouchDB \cite{anderson2010couchdb} &OrientDB \footnote{http://orientdb.com/}\\
\hline
MemcacheDB \cite{fitzpatrick2004distributed} &HyperTable \footnote{http://www.hypertable.org/} & &Pregel \cite{malewicz2010pregel}\\
\hline
Scalaris  \cite{schutt2008scalaris} & C-store \cite{StonebrakerABCCFLLMOORTZ05} & &FlockDB \footnote{https://github.com/twitter-archive/flockdb}\\
\hline
TiKV \footnote{https://www.pingcap.com/} & & &\\
\hline
\end{tabular}
\caption{\label{popstore}\footnotesize{Popular big data storage system.}}
\end{table}
\end{savenotes}

%COMPUTATION
\subsection{Data Computation}
Data computation can be from simple SQL-like query to complex machine learning techniques. There is no clear boundary to distinguish them, but we introduce them by two coarse categories as data query and data analysis. 

\subsubsection{Data Query}

MapReduce \cite{dean2008mapreduce}, Dryad \cite{isard2007dryad}, All-Pairs, Pregel \cite{malewicz2010pregel}, Spark \cite{zaharia2010spark} are the popular programming models and execution engines. More in-memory databases are proposed to accelerate the computation.

%MapReduce is a programming model and an associated implementation for processing and generating large data sets with a parallel, distributed algorithm on a cluster. The Dryad Project is investigating programming models for writing parallel and distributed programs to scale from a small cluster to a large data-center. Pregel is Google's scalable and fault-tolerant platform with an API that is sufficiently flexible to express arbitrary graph algorithms. Apache Spark is a fast and general engine for big data processing, with built-in modules for streaming, SQL, machine learning and graph processing.

\subsubsection{Data Analysis}

Data analysis can be from simple statistics to deep data mining technology. Nowadays, deep learnig has become a trend in analysis of big data, for example, MLlib \cite{meng2016mllib} by Java, Scipy, Theano, Caffe \cite{jia2014caffe} by Python, TensorFlow \cite{abadi2016tensorflow} by C++ and Torch.

%MLlib is Spark's machine learning library, focusing on learning algorithms and utilities, including classification, regression, clustering, collaborative filtering, dimensionality reduction, as well as underlying optimization primitives. Caffe is a deep learning framework made with expression, speed, and modularity in mind. TensorFlow is an open source software library for machine learning in various kinds of perceptual and language understanding tasks.

%CONTROL
\subsection{Data Control}
According to CAP theorem, it is not feasile for BDMS to fulfill ACID properties from traditional databases to guarantee both consistency and availability in a partition-prone distributed system. However, BASE theorem provides an alternative. Most NoSQL database system architectures favor one factor over the other.

\subsubsection{Data Transaction}
In databases, a transaction is a set of separate actions that must all be completely processed, or none processed at all. In partitioned databases, trading some consistency for availability can lead to dramatic improvements in scalability. NewSQL \cite{stonebraker2012newsql} is a class of modern relational database management systems that seek to provide the same scalable performance of NoSQL systems for OLTP read-write workloads while still maintaining the ACID guarantees of a traditional database system. Examples are TiDB, VoltDB  \cite{StonebrakerW13}, NuoDB \footnote{http://www.nuodb.com/} and Spanner \cite{Corbett2013Spanner}.

\subsubsection{Resource Management and Coordination}
Big data computation always runs on thousands of machines, which needs the resource management among clusters \cite{Zaharia2011The}. Mesos \cite{hindman2011mesos} and Hadoop YARN \cite{vavilapalli2013apache} are two popular resource management systems, and Apache ZooKeeper \cite{hunt2010zookeeper} enables highly reliable distributed coordination.

\subsubsection{Data Recovery}
Big data applications must be supported by a robust and rapid recovery process. The scale-out nature of the architecture can also be difficult for traditional backup applications to handle.  As database architecture has fundamentally changed to meet new application requirements, data protection needs to be redefined and re-architected as well. 

%OUTPUT
\subsection{Data Output}\label{output}
Visualization is the best way to present the results of big data management. Besides,  in this section, we also mention data sharing as a way of data output.

\subsubsection{Data Visualization}
Visualization helps us take a deep look into the big data, which provides us a interactive and graphic way to embrase the inside of big data. Tools like Tableau\footnote{http://www.tableau.com/}, Plotly\footnote{https://plot.ly/}, Visual.ly\footnote{http://visual.ly/}, Zeppelin\footnote{https://zeppelin.apache.org/} are emerging. Scalability and dynamics are two major challenges \cite{wang2015big}.

\subsubsection{Data Sharing}
Sharing data \cite{poldrack2014making} can increase the potential benefit to society of the subject's participation by providing greater opportunities for scientific discovery, researchers may have an ethical duty to share their data unless doing so would increase risk to the subjects. However, some guidelines our regulations should be made to lead us properly share data, rather than share all or share nothing.

\section{Build Up Datar}\label{bud}

\subsection{Datar Hypothesis}
We propose a unified framework of big data management systems, \textbf{datar}, to manage big data. The idea essentially comes from computer. A datar is a set of coherent softewares/systems that can manage big data pluggablly, automatically and intelligently with specific functionalities. Pluggibility means any of the five parts can be replaced by corresponding existing engines easily, automation means the flow from data input to data output can be executed coherently in a pipeline mode and intelligence means datar can mine valuable information in depth as well as self-driving system management.

\subsection{A Unified Framework}

Apache Beam \cite{Whittle2015The} is an advanced unified programming model and implements batch and streaming data processing jobs that run on any execution engine. BDAS, the Berkeley Data Analytics Stack, is an open source software stack that integrates software components being built by the AMPLab to make sense of big data. The core component Spark is a unified engine for big data processing. Apache AsterixDB \cite{alsubaiee2014asterixdb} is a scalable, open source BDMS, which aims to achieve ``One Size Fits A Bunch''. TiDB is a Hybrid Transactional/Analytical Processing (HTAP) database, inspired by the design of Google F1 and Google Spanner. Xiaomi Inc proposes their own open source ecosystems \cite{xiaomilei} for collecting and processing large-scale data in face of varied business requirements. 

The need for high-level abstraction in data centers is necessary \cite{Schwarzkopf15operatingsystem,Ionel2018flex}. The above systems have their characteristics to process big data as BDMSs, while datar is designed to overcome their drawbacks. Apache Beam aims to process batch and streaming data, and an abstraction of Runners is proposed to adapt different underlying runners. Datar is more specific on the unified framework, and divide the pluggable engines into five types of core components. The engines are chained to easily build your own BDMS to run job pipelines. RDD works as the abstraction data model in Spark and BigData sits the same position in datar. The difference is that Spark is developed from scratch while datar can make popular engines fit in its unified framework. AsterixDB is a full-function BDMS with strongly coupled components,  which is not easy to be compatible with other systems inherently. On the contrary, Xiaomi's platform is too loosely coupled which lacks of unified design. It is hard to reason about the data consistency, scalability and fault-tolerance offered by an assembly `gluing' together different systems. To be summarized, datar is designed in a high-level abstraction, and provides a unified framework to build your own BDMS with pluggable engines. Job pipelines are  executed on the chained engines with data model BigData automatically.

\begin{figure}
\centering
\includegraphics[width=1.0\columnwidth]{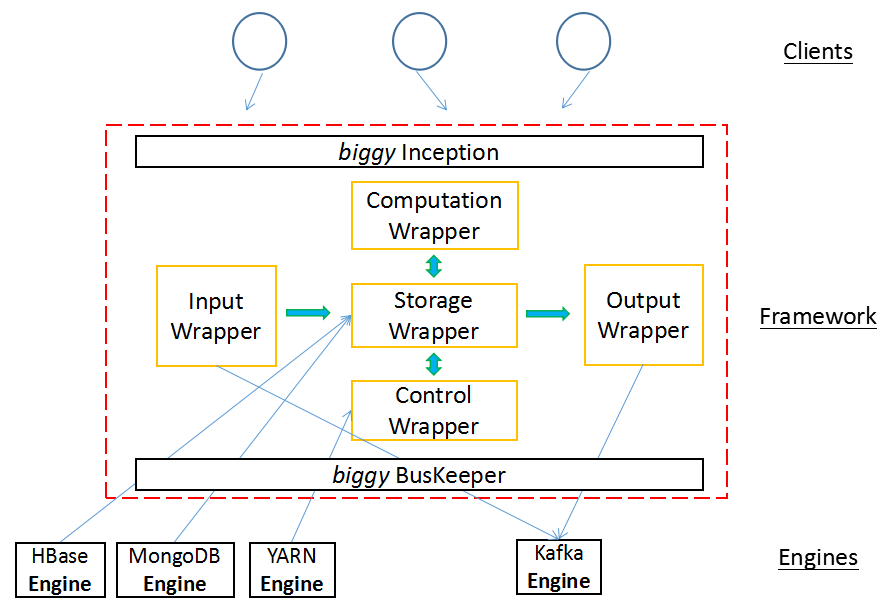}
\centering\caption{\label{fradatar}\footnotesize{Framework of datar.}}
\end{figure}

Fig. \ref{fradatar} shows the unified framework of datar, which consists of three main parts, Clients, Framework and Engines. Clients are interactive interfaces for users to interact with the running instance. Framework is the core of datar, which makes the engines pluggable, dataflow automatic and management intelligent. Engines are plugged to the unified framework. Different Wrappers need to be implemented to adapt corresponding engines into biggy Framework. In Framework part, biggy Inception parses the requests from clients and biggy BusKeeper collects the context information of running instance to help intelligent self-driving management.

\subsection{Key Features and Techniques}
Datar provides a unified framework to build a customized BDMS that can manage big data pluggablly, automatically and intelligently. We present the key techniques that achieve these features.

\subsubsection{Configuration Chain}
Users can build their own BDMS based on biggy framework by ConfChain. The configuration chain is connected by pluggable engines of five types. Once users configure the underlying engines by editing configuration file in biggy, the biggy instance generates a particular BDMS based on the unified framework according to users' configurations of engines.

\subsubsection{Data Flow}
We implement a new data structure BigData, which supports fault-tolarence with linage techniques and is suitable for distributed computation. BigData works similarly as RDD does, which is designed for our biggy data state management. BigData can transform with HDFS data, File data, HBase data by hdfsBD, fileBD, hbaseDB. BigData supports two types of operators, Action and Transformation. An Action operator creates a new BigData and a Transformation operator updates data within previous BigData. Linage techniques keep the changes of BigData during all operations. 

\subsubsection{Job Pipeline}
Jobs run on biggy within BigData by Pipeline. There are five raw Pipes including InputPipe, StoragePipe, ControlPipe, ComputationPipe and OutputPipe. A Job Pipeline consists of several Pipes, and Pipes are connected in a pipeline mode. Each Pipe includes a set of tasks belonging to a certain class, for example, a StoragePipe may contain WriteToHBaseTask and WriteToFileTask. Tasks are the actual execution units in a job pipeline.

\subsubsection{Intelligent Management and Interactive Clients}
biggy collects context information to achieve self-driving intelligent management. We have not implemented the intelligent features in our released gamma-version, but some related work can be found on this direction \cite{HellersteinSGSA17,PavloAALLMMMPQS17}. We provide interactive clients for user to access the BDMS to manipulate, including desktop, web and command line. Currently, basic command-line manipulations are supported.

\section{Play with biggy}\label{play}

\begin{figure}
\centering
\includegraphics[width=1.0\columnwidth]{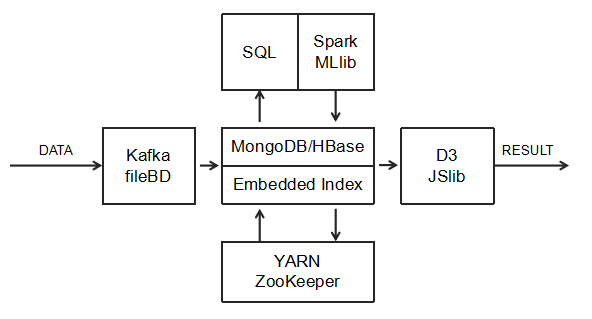}
\centering\caption{\label{imp}\footnotesize{Implementation of biggy.}}
\end{figure}

\subsection{biggy Implementation}
To put the envisioned datar into practice, we implement it as \textbf{biggy}. The greatest difficulty in implementing biggy is how to make abstractions, high-level abstractions for unified framework and low-level abstractions for flexible functionalities. biggy is based on several different system engines to fulfill the functions of input, storage, computation, control and output. Fig. \ref{imp} shows the implementation framework of biggy. Currently, we implement biggy based on Kafka, MongoDB/HBase, Spark MLlib, YARN/ZooKeeper, and D3. Kafka reads data from external sources into biggy. fildBD is the function we implement to read files on disk to biggy BigData data model. MongoDB and HBase are the two supporting storage engines. Indexes are the internal ones in them. The released gamma-version of biggy is a standalone version, but we still plug YARN as resources manager and ZooKeeper as coordination service to make biggy full-functional. For data computation, MongoDB/HBase SQL is used for basic data query by drivers, and Spark MLlib for complex data analysis. Data output takes D3 JSlib as the pluggable visualization tool to show the results on web pages. 

By different levels of abstractions and implementation, we make biggy more pluggable and automatic rather than just gluing them together. Further work of supporting most popular systems (e.g., TensorFlow) as plugins needs done with fulfillment of intelligence. Current biggy is a standalone implementation of datar by different popular engines that can run job pipelines with BigData model.

\subsection{Code Organization} 
\begin{figure}
\centering
\includegraphics[width=1.0\columnwidth]{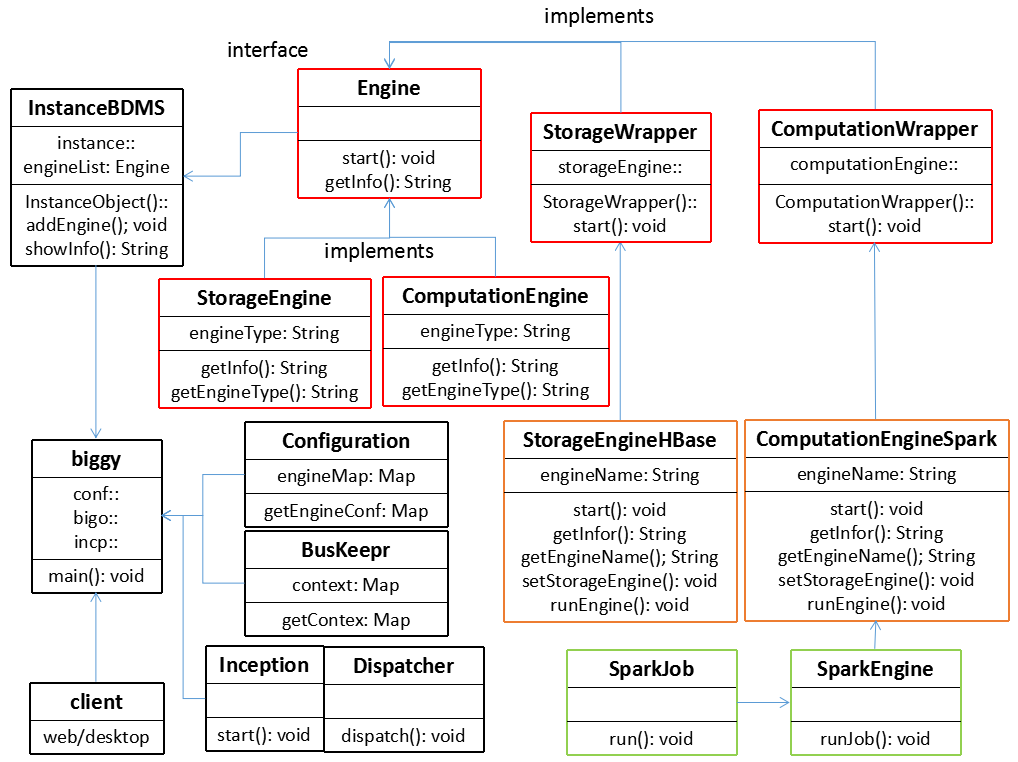}
\centering\caption{\label{codeorg}\footnotesize{Code organization og biggy.}}
\end{figure}

We implement biggy in Java by Maven as the project management tool. Several design patterns are used to make the code work such as Factory Pattern, Singleton Pattern and Chain of Responsibility Pattern. The simplified overview of code organization is shown in Fig. \ref{codeorg}. A high-resolution picture can be found here\footnote{https://www.processon.com/view/link/5b0e4eeee4b06350d445fcb3}. As we can see, biggy instance bigo sits at the center, the pluggability on the upper side is based on factory pattern and chained together, the automation is based on job pipelines and BigData model at the lower side, the clients are left-side and intelligent part is right-side. Currently, the clients and intelligent parts are unimplemented, and biggy is set on standalone mode.

\begin{figure}
\centering
\includegraphics[width=1.0\columnwidth]{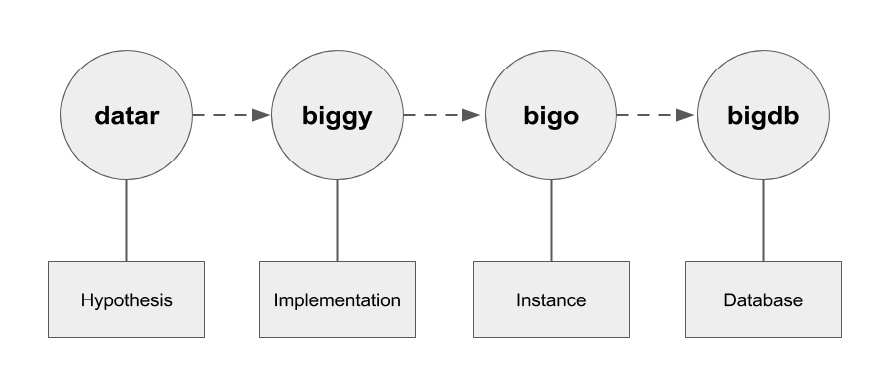}
\centering\caption{\label{relation}\footnotesize{Relation among several key terms.}}
\end{figure}

\begin{figure}
\centering
\includegraphics[width=0.8\columnwidth]{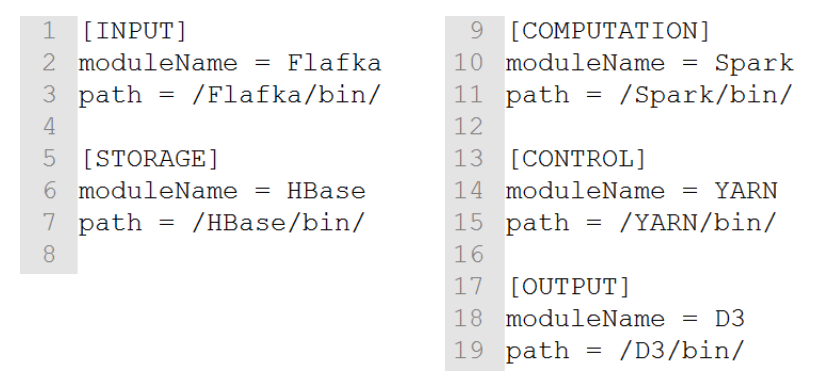}
\centering\caption{\label{conf}\footnotesize{Configuration of biggy.}}
\end{figure}

\begin{figure}
\centering
\includegraphics[width=1.0\columnwidth]{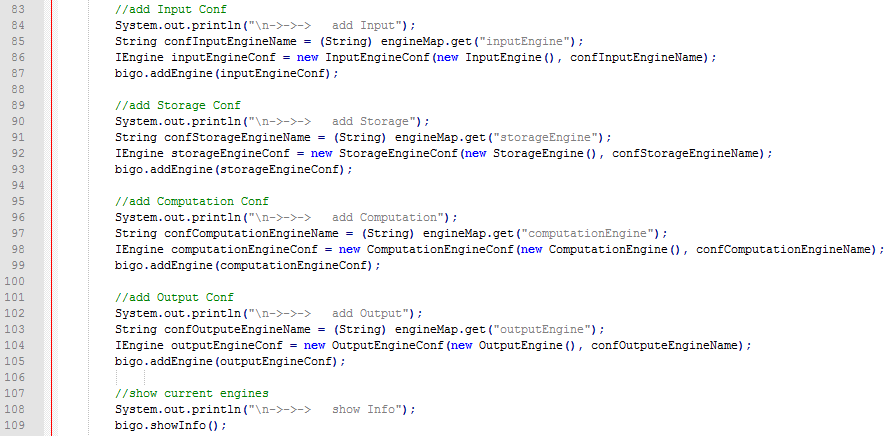}
\centering\caption{\label{code}\footnotesize{Example of biggy ConfChain.}}
\end{figure}

Code in black color is the core of biggy, code in red is the main framework of biggy, code in dark orange depends on the pluggable engines, and code in green is application-specific. To see implementation details please check out the source code with gamma-version.

\subsection{Opensource Development}
There are three types of users for biggy, i.e., Engine Writers, Framework Writers and App (-lication) Writers. The project is opensourced and can be found at Github\footnote{https://github.com/Ideamaxwu/biggy}. We focus on the implementation of Framework to provide a unified framework for customizing your own BDMS. Engine Writers are responsible to develop new pluggable engines (e.g., HBase) and adapt them into biggy framework by Wrappers. Some Wrapper templates are provided for Engine Writers to use. App Writers are the  end users of biggy, who can run their application-specific jobs on a customized BDMS.

\subsection{biggy Deployment}

For better explanation, relations among several key terms are shown in Fig. \ref{relation}. bigo is an instance of biggy and bigdb is the physical files where store the data. Fig. \ref{conf} shows the configuration of biggy. Before installation and use of biggy, it is necessary to set the five parts in configuration file. biggy checks if the targeted engines are available. For each module, there should be a Wrapper to make it fit in biggy, which is implemented by programmers and can be shared for public use. That is why we make it opensourced and any modules/systems/libraries follow the design of biggy can be plugged in to work. Fig. \ref{code} shows how to chain the pluggable engines, in which bigo is the running instance of biggy. Servers of the pluggable engines should get started, such as MongoDB server and Kafka Server. After deployment of a customized BDMS, App Writers can run job pipelines on it with easy coding.

\subsection{biggy Demonstrations}

\begin{figure}
\subfigure[WordCount example]{
    \includegraphics[width=1.0\columnwidth]{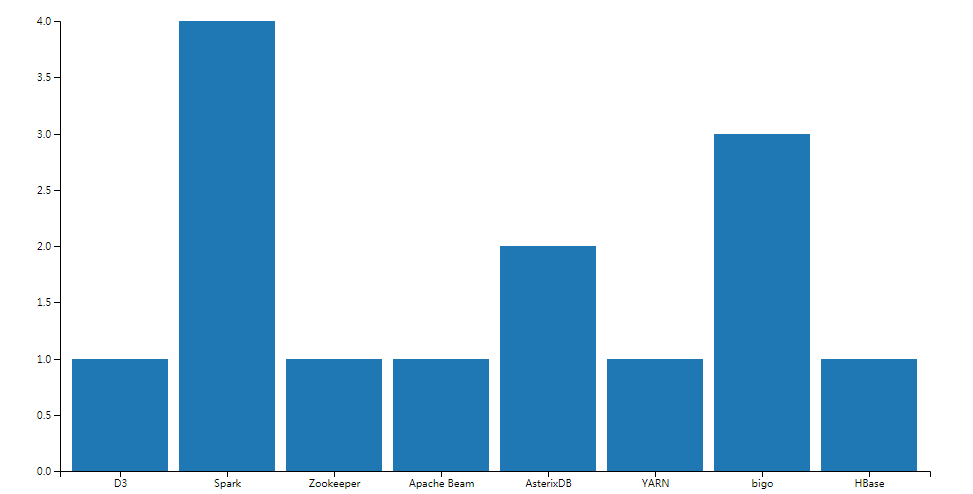}}
\subfigure[Sort example]{
    \includegraphics[width=1.0\columnwidth]{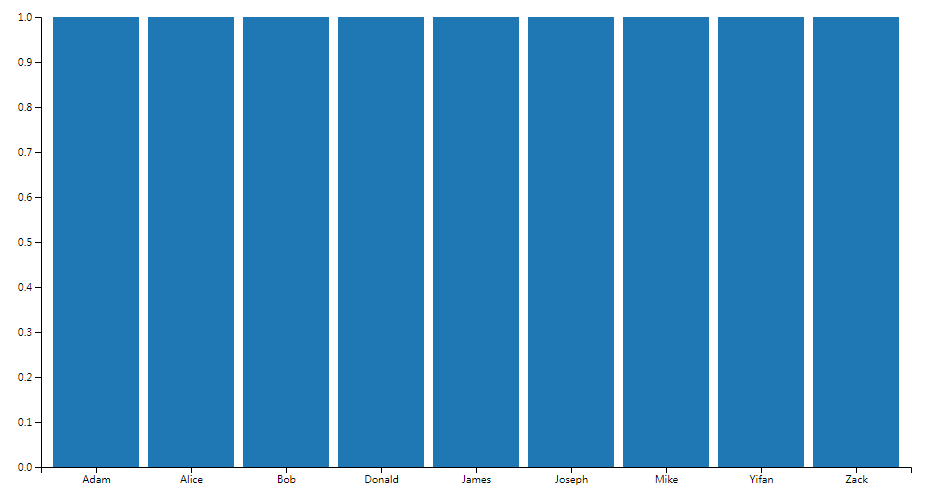}}
\subfigure[KMeans example]{
    \includegraphics[width=0.7\columnwidth]{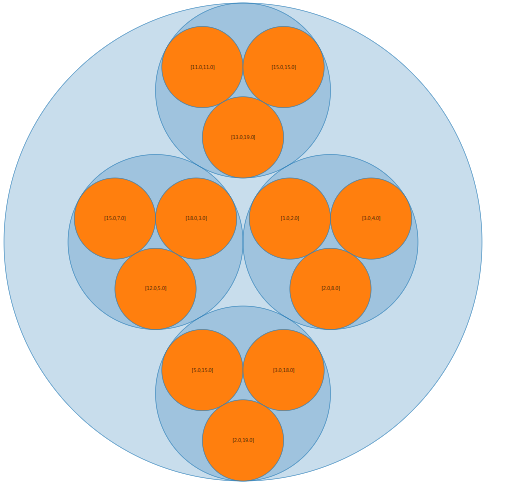}}
\subfigure[PageRank example]{
    \includegraphics[width=1.0\columnwidth]{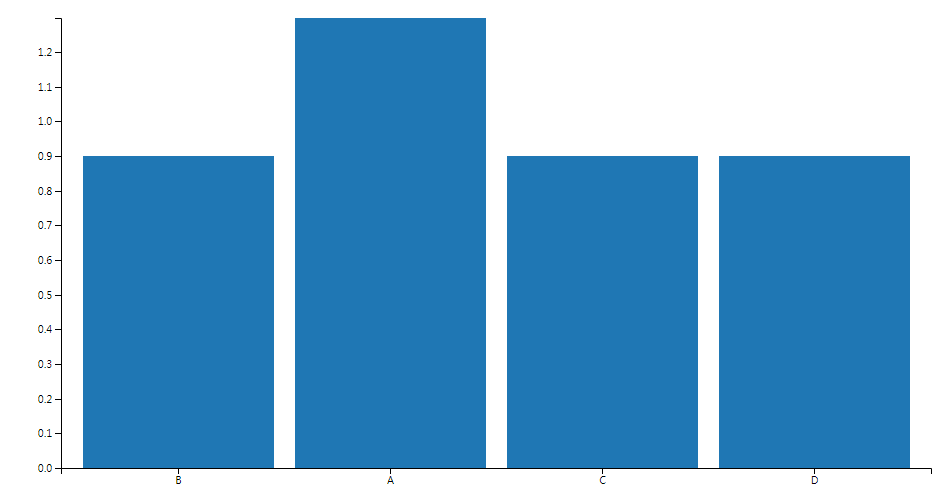}}
\centering\caption{\label{mani}\footnotesize{Examples of biggy demonstrations.}}
\end{figure}

We present four running examples to show how biggy works. Fig. \ref{mani} shows the demonstration of biggy, including running examples WordCount, Sort, KMeans and PageRank. Take WordCount for example, given a set of words, the goal is to count the number of each word. We input data from test file \emph{egDBcount.txt} by implementing \emph{egBDIOPipeJobReadFile.java}, store the data into MongoDB by \emph{egHBasePipeJobWriteDB.java}, compute the results using Spark by \emph{egSparkPipeJobWordCount.java}, and output the visualization by \emph{egD3PipeJobVisual.java}. Control part is the default setting on a standalone instance. Two more classes of \emph{egSparkPipeJobWordCountTask001.java} and \emph{egD3PipeJobVisualTask001.java} should be concreted to realize specific data operations. Give the customized BDMS, WordCount pipeline can be executed automatically on BigData model by connecting the pipes in \emph{egWordCount.java}. The other three examples follow the similar implementation. Given the unified framework and simple APIs to use, App Writers can easily deploy their own BDMSs, run job pipelines and care little about the underlying complexity. We give a short summary for each example.

\subsubsection{Example WordCount}
Given a set of words, the goal is to count the number of each word. Fig. \ref{mani} (a) shows the results. As we can see, Spark counts four and YANR counts one.

\subsubsection{Example Sort}
Given a random set of strings, the goal is the sort them by alphabet order. In Fig. \ref{mani} (b), horizontal label shows the order from A to Z.

\subsubsection{Example KMeans}
KMeans clustering aims to partition $n$ observations into $k$ clusters in which each observation belongs to the cluster with the nearest mean. In our example, each observation is a coordinate with $(x,y)$. Fig. \ref{mani} (c) shows the clustering results.

\subsubsection{Example PageRank}
PageRank is a way of measuring the importance of website pages. In our test data, each item follows format (A $->$ B), which means A is linked to B. The importance score of each page is showed in  Fig. \ref{mani} (d).

%EVEL
\section{biggy Evaluation}\label{eval}
We evaluate biggy on examples of WordCount, Sort, KMeans, PageRank by efficiency and scalability in five parts. They are four typical classes of algorithms \cite{Shi2015Clash}. All the four algorithms are executed on Spark runtime engine as the computation engine.

\subsection{Experiment Setup}
The set of experimental evaluation is designed for standalone biggy on a small-scale and personal-computer testbed. The performance mainly depends on the underlying pluggable engines and more experiments in real applications are welcome. From our experiments, we can get a general qualitative and quantitative understanding of biggy performance.

\subsubsection{Hardware Configuration}
The implementation of unified framework biggy is deployed on a laptop with a standalone mode. The laptop has a memory of 8GB, 64-bit Windows 7 system, Intel(R) Core(TM) i7 CPU at 2.5 GHz. We run each experiment three times and report the average results.

\subsubsection{Software Configuration}
For the set of experiments, Java version is 1.8.0, MongoDB is 3.2.3, Spark is 2.3.0, D3 is version 4 and biggy version is gamma.

\subsubsection{Data Preparation}
We use different datasets for four examples, and the dataset samples can be found in the source code. We generate random data from 1K to 1M to evaluate biggy on our laptop. For time and space efficiency, we use default data size 1M, and varied data size is applied for scalability evaluation. In KMeans, the number of clusters is 100 and the number of iterations is 20.  In PageRank, the default number of pages is 1,000. To be noted, 1M means the number of items is 1M, not the disk storage, and the total disk storage depends on the size of each item.

\subsection{Time Efficiency}
\begin{figure}
\centering
\includegraphics[width=1.0\columnwidth]{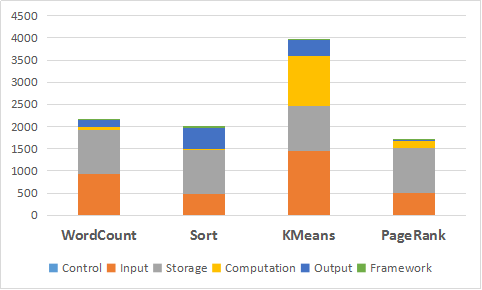}
\centering\caption{\label{evel00}\footnotesize{Time efficiency on different types of algorithms.}}
\end{figure}

Time efficiency is shown in Fig. \ref{evel00}. Because Storage and Computation take too much time in real applications, for better display effects, we count these two parts in seconds and other parts in milliseconds. As we can see, storage is the most time-consuming part, and the reason is we store the data into MongoDB by inserting every tuple, which costs a lot. Computation comes later and Input/Output do not take much time. Control part takes little time because we run experiments in standalone mode. The negligible time of Framework shows the efficiency of our proposed unified framework. KMeans ranks top due to iterations during clustering.

\subsection{Space Efficiency}
\begin{figure}
\centering
\includegraphics[width=1.0\columnwidth]{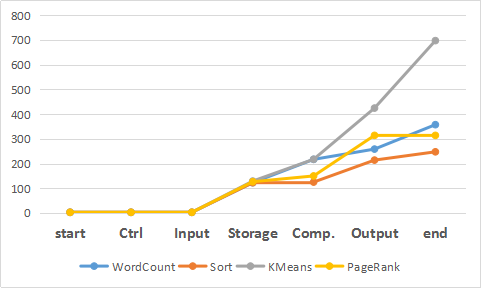}
\centering\caption{\label{evel01}\footnotesize{Space efficiency on different types of algorithms.}}
\end{figure}

Space efficiency is shown in Fig. \ref{evel01}. We evaluate used memory of each algorithm, which is calculated by the difference of total amount of memory and amount of free memory in the Java virtual machine. And the used memory is measured in MBs. Fig.  \ref{evel01} shows the usage of memory in Java VM at the start of each Pipe stage. As we can see, used memory increases as the pipeline goes on and KMeans costs most, too. 

\subsection{Scalability}
\begin{figure}
\centering
\includegraphics[width=1.0\columnwidth]{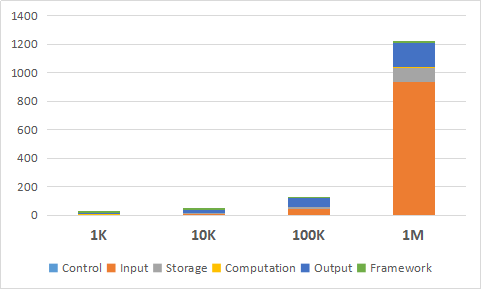}
\centering\caption{\label{evel02}\footnotesize{Scalability on different size of datasets for time efficiency.}}
\end{figure}

\begin{figure}
\centering
\includegraphics[width=1.0\columnwidth]{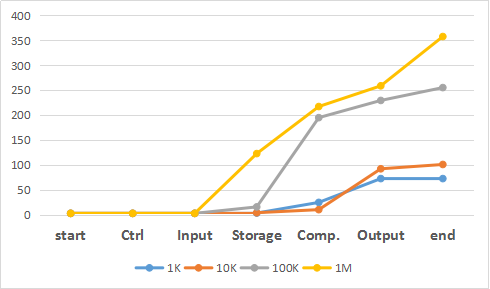}
\centering\caption{\label{evel03}\footnotesize{Scalability on different size of datasets for space efficiency.}}
\end{figure}

Scalability evaluation is shown in Fig. \ref{evel02} and Fig. \ref{evel03} and is based on WordCount example. Table \ref{sdetail} shows the details of the results on scalability for time efficiency. The data size varies from 1K to 1M, both time and space increases with the size of data, which is reasonable. Same as previous experiments, Storage and Computation time is measured in seconds, and the rest in milliseconds. We do not optimize the garbage collection of Java in this released version. If so, better results may get. For limitation of pages, detailed results are on Github. % in folder \emph{examples/Exps} 

\begin{table}
\centering
\begin{tabular}{|c|c|c|c|c|c|c|}
\hline
\textbf{Data Size} &1K &10K &100K &1M \\
\hline
\hline
\textbf{Control} &0 &1 &1 &1\\
\hline
\textbf{Input} &3 &9 &42 &933\\
\hline
\textbf{Storage} &1.809 (s) &3.337 (s) &12.797 (s) &98.91 (s)\\
\hline
\textbf{Computation} &2.859 (s) &3.077 (s) &3.367 (s) &5.766 (s)\\
\hline
\textbf{Output} &8 &22 &59 &173\\
\hline
\textbf{Framework} &5 &6 &6 &4\\
\hline
\end{tabular}
\caption{\label{sdetail}\footnotesize{Details of scalability for time efficiency.}}
\end{table}

%CONC
\section{Conclusion And Future Work}\label{conc}
We propose datar as a unified framework for BDMS from the perspective of a computer in five components and present biggy as an implementation of datar to manage big data pluggablly, automatically and intelligently with specific functionalities. biggy presents only the first step towards a unified framework of big data management system, and we hope it helps frame important research questions. In particular, some progress we are working on include:

\begin{itemize}
\item Implementation of intelligent self-driving management .
\item From standalone to cluster mode.
\item System optimization to improve efficiency.
\item More popular engines support and multi-OS support.
\end{itemize}

Many technical challenges must be addressed before this potential unified framework can be fully realized. We encourage fundamental research towards addressing the architecture design and technical challenges to achieve the promised benefits of big data. More resources about biggy are accessible on Github and so is the contact information.

%bib
\small
\bibliographystyle{IEEEtran}
\input{BDMS_ICPAD2018_8p.bbl}

\end{document}

%% file: BDMS_ICPAD2018_8p.bbl

%% file: BDMS_ICPAD2018_8p.bbl
\begin{thebibliography}{10}
\providecommand{\url}[1]{#1}
\csname url@samestyle\endcsname
\providecommand{\newblock}{\relax}
\providecommand{\bibinfo}[2]{#2}
\providecommand{\BIBentrySTDinterwordspacing}{\spaceskip=0pt\relax}
\providecommand{\BIBentryALTinterwordstretchfactor}{4}
\providecommand{\BIBentryALTinterwordspacing}{\spaceskip=\fontdimen2\font plus
\BIBentryALTinterwordstretchfactor\fontdimen3\font minus
  \fontdimen4\font\relax}
\providecommand{\BIBforeignlanguage}[2]{{%
\expandafter\ifx\csname l@#1\endcsname\relax
\typeout{** WARNING: IEEEtran.bst: No hyphenation pattern has been}%
\typeout{** loaded for the language `#1'. Using the pattern for}%
\typeout{** the default language instead.}%
\else
\language=\csname l@#1\endcsname
\fi
#2}}
\providecommand{\BIBdecl}{\relax}
\BIBdecl

\bibitem{turing1950computing}
A.~M. Turing, ``Computing machinery and intelligence,'' \emph{Mind}, vol.~59,
  no. 236, pp. 433--460, 1950.

\bibitem{von1993first}
J.~Von~Neumann, ``First draft of a report on the edvac,'' \emph{IEEE Annals of
  the History of Computing}, no.~4, pp. 27--75, 1993.

\bibitem{Bachman66}
C.~W. Bachman, ``On a generalized language for file organization and
  manipulation,'' \emph{Commun. {ACM}}, vol.~9, no.~3, pp. 225--226, 1966.

\bibitem{Codd70}
E.~F. Codd, ``A relational model of data for large shared data banks,''
  \emph{Commun. {ACM}}, vol.~13, no.~6, pp. 377--387, 1970.

\bibitem{StonebrakerWKH76}
M.~Stonebraker, E.~Wong, P.~Kreps, and G.~Held, ``The design and implementation
  of {INGRES},'' \emph{{ACM} Trans. Database Syst.}, vol.~1, no.~3, pp.
  189--222, 1976.

\bibitem{Stonebraker90}
M.~Stonebraker, ``The postgres {DBMS},'' in \emph{Proceedings of the {ACM}
  {SIGMOD}}, 1990, p. 394.

\bibitem{StonebrakerALPSSSY96}
M.~Stonebraker, P.~M. Aoki, W.~Litwin, A.~Pfeffer, A.~Sah, J.~Sidell,
  C.~Staelin, and A.~Yu, ``Mariposa: {A} wide-area distributed database
  system,'' \emph{{VLDB} J.}, vol.~5, no.~1, pp. 48--63, 1996.

\bibitem{StonebrakerABCCFLLMOORTZ05}
M.~Stonebraker, D.~J. Abadi, A.~Batkin, X.~Chen, M.~Cherniack, and et~al,
  ``C-store: {A} column-oriented {DBMS},'' in \emph{Proceedings of VLDB}, 2005,
  pp. 553--564.

\bibitem{StonebrakerW13}
M.~Stonebraker and A.~Weisberg, ``The voltdb main memory {DBMS},'' \emph{{IEEE}
  Data Eng. Bull.}, vol.~36, no.~2, pp. 21--27, 2013.

\bibitem{alsubaiee2014asterixdb}
S.~Alsubaiee, Y.~Altowim, H.~Altwaijry, A.~Behm \emph{et~al.}, ``Asterixdb: A
  scalable, open source bdms,'' \emph{Proceedings of the VLDB Endowment},
  vol.~7, no.~14, pp. 1905--1916, 2014.

\bibitem{TaftESLASMA18}
R.~Taft, N.~El{-}Sayed, M.~Serafini, Y.~Lu, A.~Aboulnaga, M.~Stonebraker,
  R.~Mayerhofer, and F.~Andrade, ``P-store: An elastic database system with
  predictive provisioning,'' in \emph{Proceedings of the 2018 SIGMOD}, 2018,
  pp. 205--219.

\bibitem{Baker2011Megastore}
J.~Baker, ``Megastore : Providing scalable, highly available storage for
  interactive services,'' in \emph{Biennial Conference on Innovative Data
  Systems Research}, 2011, pp. 223--234.

\bibitem{Corbett2013Spanner}
J.~C. Corbett, J.~Dean, M.~Epstein, A.~Fikes, and et~al, ``Spanner: Google's
  globally-distributed database,'' \emph{Acm Transactions on Computer Systems},
  vol.~31, no.~3, p.~8, 2013.

\bibitem{Akidau2013MillWheel}
T.~Akidau, A.~Balikov, K.~Bekirolu, S.~Chernyak \emph{et~al.}, ``Millwheel:
  fault-tolerant stream processing at internet scale,'' \emph{Proceedings of
  the VLDB Endowment}, vol.~6, no.~11, pp. 1033--1044, 2013.

\bibitem{Hey2012The}
T.~Hey, ``The fourth paradigm ¨c data-intensive scientific discovery,'' in
  \emph{International Symposium on Information Management in a Changing World},
  2012, pp. 1--1.

\bibitem{grover2015data}
R.~Grover and M.~J. Carey, ``Data ingestion in asterixdb.'' in \emph{EDBT},
  2015, pp. 605--616.

\bibitem{alsubaiee2014storage}
S.~Alsubaiee, A.~Behm, V.~Borkar, Z.~Heilbron, Y.-S. Kim, M.~J. Carey,
  M.~Dreseler, and C.~Li, ``Storage management in asterixdb,''
  \emph{Proceedings of the VLDB Endowment}, vol.~7, no.~10, pp. 841--852, 2014.

\bibitem{Bu2016Large}
Y.~Bu, Y.~Bu, Y.~Bu, Y.~Bu, and Y.~Bu, ``Large-scale complex analytics on
  semi-structured datasets using asterixdb and spark,'' \emph{Proceedings of
  the VLDB Endowment}, vol.~9, no.~13, pp. 1585--1588, 2016.

\bibitem{borkar2011hyracks}
V.~Borkar, M.~Carey, R.~Grover, N.~Onose, and R.~Vernica, ``Hyracks: A flexible
  and extensible foundation for data-intensive computing,'' in \emph{2011 IEEE
  27th ICDE}.\hskip 1em plus 0.5em minus 0.4em\relax IEEE, 2011, pp.
  1151--1162.

\bibitem{adamu2015survey}
F.~B. Adamu, A.~Habbal, S.~Hassan, R.~Les~Cottrell, B.~White, and I.~Abdullahi,
  ``A survey on big data indexing strategies.''

\bibitem{decandia2007dynamo}
G.~DeCandia, D.~Hastorun, M.~Jampani, G.~Kakulapati, and et~al, ``Dynamo:
  amazon's highly available key-value store,'' \emph{ACM SIGOPS Operating
  Systems Review}, vol.~41, no.~6, pp. 205--220, 2007.

\bibitem{chang2008bigtable}
F.~Chang, J.~Dean, S.~Ghemawat, W.~C. Hsieh, and et~al, ``Bigtable: A
  distributed storage system for structured data,'' \emph{ACM Transactions on
  Computer Systems}, vol.~26, no.~2, p.~4, 2008.

\bibitem{banker2011mongodb}
K.~Banker, \emph{MongoDB in action}.\hskip 1em plus 0.5em minus 0.4em\relax
  Manning Publications Co., 2011.

\bibitem{sumbaly2012serving}
R.~Sumbaly, J.~Kreps, L.~Gao, A.~Feinberg, C.~Soman, and S.~Shah, ``Serving
  large-scale batch computed data with project voldemort,'' in
  \emph{Proceedings of the 10th USENIX conference on File and Storage
  Technologies}.\hskip 1em plus 0.5em minus 0.4em\relax USENIX Association,
  2012, pp. 18--18.

\bibitem{lakshman2010cassandra}
A.~Lakshman and P.~Malik, ``Cassandra: a decentralized structured storage
  system,'' \emph{ACM SIGOPS Operating Systems Review}, vol.~44, no.~2, pp.
  35--40, 2010.

\bibitem{carlson2013redis}
J.~L. Carlson, \emph{Redis in Action}.\hskip 1em plus 0.5em minus 0.4em\relax
  Manning Publications Co., 2013.

\bibitem{george2011hbase}
L.~George, \emph{HBase: the definitive guide}.\hskip 1em plus 0.5em minus
  0.4em\relax " O'Reilly Media, Inc.", 2011.

\bibitem{anderson2010couchdb}
J.~C. Anderson, J.~Lehnardt, and N.~Slater, \emph{CouchDB: the definitive
  guide}.\hskip 1em plus 0.5em minus 0.4em\relax " O'Reilly Media, Inc.", 2010.

\bibitem{fitzpatrick2004distributed}
B.~Fitzpatrick, ``Distributed caching with memcached,'' \emph{Linux journal},
  vol. 2004, no. 124, p.~5, 2004.

\bibitem{malewicz2010pregel}
G.~Malewicz, M.~H. Austern, A.~J. Bik, and et~al, ``Pregel: a system for
  large-scale graph processing,'' in \emph{Proceedings of the 2010 ACM
  SIGMOD}.\hskip 1em plus 0.5em minus 0.4em\relax ACM, 2010, pp. 135--146.

\bibitem{schutt2008scalaris}
T.~Sch{\"u}tt, F.~Schintke, and A.~Reinefeld, ``Scalaris: reliable
  transactional p2p key/value store,'' in \emph{Proceedings of the 7th ACM
  SIGPLAN workshop on ERLANG}.\hskip 1em plus 0.5em minus 0.4em\relax ACM,
  2008, pp. 41--48.

\bibitem{dean2008mapreduce}
J.~Dean and S.~Ghemawat, ``Mapreduce: simplified data processing on large
  clusters,'' \emph{Communications of the ACM}, vol.~51, no.~1, pp. 107--113,
  2008.

\bibitem{isard2007dryad}
M.~Isard, M.~Budiu, Y.~Yu, A.~Birrell, and D.~Fetterly, ``Dryad: distributed
  data-parallel programs from sequential building blocks,'' in \emph{ACM SIGOPS
  Operating Systems Review}, vol.~41, no.~3.\hskip 1em plus 0.5em minus
  0.4em\relax ACM, 2007, pp. 59--72.

\bibitem{zaharia2010spark}
M.~Zaharia, M.~Chowdhury, M.~J. Franklin, S.~Shenker, and I.~Stoica, ``Spark:
  cluster computing with working sets.'' \emph{HotCloud}, vol.~10, pp. 10--10,
  2010.

\bibitem{meng2016mllib}
X.~Meng, J.~Bradley, B.~Yuvaz, E.~Sparks, S.~Venkataraman \emph{et~al.},
  ``Mllib: Machine learning in apache spark,'' \emph{JMLR}, vol.~17, no.~34,
  pp. 1--7, 2016.

\bibitem{jia2014caffe}
Y.~Jia, E.~Shelhamer, J.~Donahue, S.~Karayev, and et~al, ``Caffe: Convolutional
  architecture for fast feature embedding,'' in \emph{Proceedings of the 22nd
  ACM international conference on Multimedia}.\hskip 1em plus 0.5em minus
  0.4em\relax ACM, 2014, pp. 675--678.

\bibitem{abadi2016tensorflow}
M.~Abadi, A.~Agarwal, P.~Barham, E.~Brevdo \emph{et~al.}, ``Tensorflow:
  Large-scale machine learning on heterogeneous distributed systems,''
  \emph{arXiv preprint arXiv:1603.04467}, 2016.

\bibitem{stonebraker2012newsql}
M.~Stonebraker, ``Newsql: An alternative to nosql and old sql for new oltp
  apps,'' \emph{Communications of the ACM. Retrieved}, pp. 07--06, 2012.

\bibitem{Zaharia2011The}
M.~Zaharia, B.~Hindman, A.~Konwinski, A.~Ghodsi \emph{et~al.}, ``The datacenter
  needs an operating system,'' in \emph{Usenix Conference on Hot Topics in
  Cloud Computing}, 2011, pp. 17--17.

\bibitem{hindman2011mesos}
B.~Hindman, A.~Konwinski, M.~Zaharia, A.~Ghodsi, and et~al, ``Mesos: A platform
  for fine-grained resource sharing in the data center.'' in \emph{NSDI},
  vol.~11, no. 2011, 2011, pp. 22--22.

\bibitem{vavilapalli2013apache}
V.~K. Vavilapalli, A.~C. Murthy, C.~Douglas \emph{et~al.}, ``Apache hadoop
  yarn: Yet another resource negotiator,'' in \emph{Proceedings of the 4th
  annual Symposium on Cloud Computing}.\hskip 1em plus 0.5em minus 0.4em\relax
  ACM, 2013, p.~5.

\bibitem{hunt2010zookeeper}
P.~Hunt, M.~Konar, F.~P. Junqueira, and B.~Reed, ``Zookeeper: Wait-free
  coordination for internet-scale systems.'' in \emph{USENIX annual technical
  conference}, vol.~8, 2010, p.~9.

\bibitem{wang2015big}
L.~Wang, G.~Wang, and C.~A. Alexander, ``Big data and visualization: methods,
  challenges and technology progress,'' \emph{Digital Technologies}, vol.~1,
  no.~1, pp. 33--38, 2015.

\bibitem{poldrack2014making}
R.~A. Poldrack and K.~J. Gorgolewski, ``Making big data open: data sharing in
  neuroimaging,'' \emph{Nature neuroscience}, vol.~17, no.~11, pp. 1510--1517,
  2014.

\bibitem{Whittle2015The}
T.~Akidau, R.~Bradshaw, C.~Chambers, S.~Chernyak,
  R.~Fern{\'{a}}ndez{-}Moctezuma, R.~Lax, S.~McVeety, D.~Mills, F.~Perry,
  E.~Schmidt, and S.~Whittle, ``The dataflow model: a practical approach to
  balancing correctness, latency, and cost in massive-scale, unbounded,
  out-of-order data processing,'' \emph{Proceedings of the Vldb Endowment},
  vol.~8, no.~12, pp. 1792--1803, 2015.

\bibitem{xiaomilei}
L.~Jun, Y.~Hangjun, W.~Zesheng, Z.~Peng, X.~Long, and H.~Yanxiang, ``Big-data
  platform based on open source ecosystem,'' \emph{Journal of Computer Research
  and Development}, vol.~54, no.~1, pp. 80--93, 2017.

\bibitem{Schwarzkopf15operatingsystem}
M.~Schwarzkopf, ``Operating system support for warehouse-scale computing,''
  2015.

\bibitem{Ionel2018flex}
I.~Gog, ``Flexible and efficient computation in large data centres,'' 2018.

\bibitem{HellersteinSGSA17}
J.~M. Hellerstein, V.~Sreekanti, J.~E. Gonzalez, J.~Dalton, A.~Dey, S.~Nag, and
  et~al, ``Ground: {A} data context service,'' in \emph{{CIDR} 2017, 8th
  Biennial Conference on Innovative Data Systems Research}, 2017.

\bibitem{PavloAALLMMMPQS17}
A.~Pavlo, G.~Angulo, J.~Arulraj, H.~Lin, J.~Lin, L.~Ma, and et~al,
  ``Self-driving database management systems,'' in \emph{{CIDR} 2017, 8th
  Biennial Conference on Innovative Data Systems Research}, 2017.

\bibitem{Shi2015Clash}
J.~Shi, Y.~Qiu, U.~F. Minhas, L.~Jiao, C.~Wang, and B.~Reinwald, ``Clash of the
  titans: Mapreduce vs. spark for large scale data analytics,''
  \emph{Proceedings of the VLDB Endowment}, vol.~8, no.~13, pp. 2110--2121,
  2015.

\end{thebibliography}
